\documentclass[a4paper,fleqn]{cas-sc}

\usepackage[numbers]{natbib}

\newcommand{\MS}[1]{\textcolor{red}{#1}}

\def\tsc#1{\csdef{#1}{\textsc{\lowercase{#1}}\xspace}}
\tsc{WGM}
\tsc{QE}

\begin{document}
\let\WriteBookmarks\relax
\def\floatpagepagefraction{1}
\def\textpagefraction{.001}

\ExplSyntaxOn
\keys_set:nn { stm / mktitle } { nologo }
\ExplSyntaxOff

\shorttitle{Thermomechanical Dissipative behaviour of Metallic Glasses}   


\title [mode = title]{Thermomechanical Dissipative behaviour of CuZr metallic glasses}  



\author[1]{Matias Sepulveda--Macias}

%

\ead{matsepmacias@gmail.com}


\affiliation[a1]{organization={Univ Lyon, INSA-Lyon, CNRS, LaMCoS, UMR 5259},
            city={Villeurbanne},
            postcode={69621}, 
            country={France.}}

\author[1]{Gergely Moln{\'a}r}
\ead{gergely.molnar@insa-lyon.fr}

\author[1,2]{Anne Tanguy}


\ead{anne.tanguy@insa-lyon.fr}


\affiliation[a2]{organization={ONERA},
            addressline={University Paris-Saclay, Chemin de la Huni\`ere, BP 80100}, 
            city={Palaiseau},
            postcode={92123}, 
            country={France.}}

\cortext[1]{Anne Tanguy Anne.Tanguy@insa-lyon.fr}
\cormark[1]



\begin{abstract}
We performed molecular dynamics simulations of Zr$_{50}$Cu$_{50}$ metallic glass samples submitted to mechanical deformation at different strain rates. The simultaneous measurements of the stress-strain curve, and of the temperature evolution during the 
cyclic mechanical load, are used to determine the thermo-mechanical constitutive laws at the continuum scale. It is shown that plastic deformation acts as a heat source, but strong finite size effects affect the unfolding of shear bands and its related dissipation rate. Finally, a thermo-mechanical constitutive law is proposed to reproduce quantitatively self-heating processes at different scales.

\end{abstract}




\maketitle

\section{Introduction}\label{Intro}
Metallic glasses (MGs) are metallic alloys that lack of intrinsic internal length scale. Their amorphous structure is obtained from a rapidly quenched melt, that is, they undergo a glass transition leading to structural properties of a liquid, with dynamical properties of a solid. This disordered nature makes them exhibit excellent mechanical properties, such as high yield strength, large elastic strain limits, good wear resistance, among other~\cite{Johnson1999}. Research in practical applications includes nanotechnology, micro electromechanical systems and bio-medical devices~\cite{Chen2008,Hufnagel2016} \MS{for which thermal management is of strong interest}. The mechanical properties of MGs, including an atomic level explanation of the mechanism of elasto-plastic regime, and the mechanism for temperature dependence of the plasticity of MGs remains however a subject of intense research~\cite{Tanguy2022,Lagogianni2022}\MS{, since it differs greatly from that involved in crystals}. It has been shown that, in the low strain rate low temperature limit for example, plastic deformation results from a succession of local Eshelby-like plastic rearrangements~\cite{Albaret2016}, while the elastic response shows nanometer-scale disorder-induced correlations, appearing as heterogeneous mechanical assemblies, responsible for the peculiar vibrational response of these amorphous materials~\cite{Tanguy2002, Beltukov2016}. It is shown that mechanical wave-packets can induce local heating and heat transfer in amorphous solids, due to scattering processes on structural disorder~\cite{Beltukov2018}. This means that thermal and mechanical response are intimately intertwined, and so may depend on the strain rate as well as on the surrounding temperature~\cite{Fusco2010,SepulvedaMacias2016,SepulvedaMacias2018}.

Understanding the thermomechanical properties is essential in materials science, engineering, and physics, since the temperature changes inside a material could have a significant impact on their strength, ductility, and other mechanical properties~\cite{Rosner2014,Mota2021}. Thermal measurement can also be used as damage warning devices, or to confirm the integrity of a system~\cite{Berthe2023}. \MS{The fraction of plastic work converted to heat during large strain mechanical deformation (Taylor–Quinney coefficient) has been studied in some crystalline materials,  either experimentally~\cite{Ravichandran2002} or with the help of Molecular Dynamics Simulations~\cite{Kositski2021,Xiong2022,Signetti2023}. The result is very sensitive to the materials structure and composition. The complete }study of thermomechanical properties involves analysing the response of materials to different temperatures and stress loadings, determining self-heating properties, and predicting the behaviour of materials using constitutive models. In this aspect, within the framework of the generalized standard materials theory (GSM), Chrysochoos and Belmahjoub~\cite{Chrysochoos1992,Peyroux1998} used the information obtained from the global mechanical laws and the thermal changes during cycles of shear, to identify thermo-mechanical constitutive laws and explain the energy balance evolution during deformation for duraluminium samples. They observed that the model based on the classical theory of the time-independent elastoplasticity gives correct mechanical predictions~\cite{Chrysochoos1992}.

Closer to our study, Zhao and Li studied the evolution of temperature in the shear banding process for bulk metallic glasses using the finite-element method \MS{with an empirical constitutive law}~\cite{Zhao2011}. They show results on tension and compression of Vitreloy 1, where the work accumulated during plastic deformation is used as a heat source, to solve the heat conduction equation, yielding to an increase in temperature \MS{of a few degrees} inside the shear band. This phenomenon of temperature increase within the shear band has also been studied \MS{at the atomistic scale} by Lagogianni and Varnik~\cite{Lagogianni2022}. The authors use non-equilibrium molecular dynamics to simulate a Lennard-Jones type binary mixture. As in the article of Zhao and Li~\cite{Zhao2011}, Lagogianni and Varnik show that the zones supporting the largest strains contribute the most to the local temperature rise, thus yielding to a strong increase in temperature within the shear band compared to the matrix outside it. 

Here we present a molecular dynamics study on Zr$_{50}$Cu$_{50}$ metallic glass subjected to cycles of shear. The global constitutive laws and the analysis of the thermal behaviour at the micro-scale are both considered \MS{with a special focus on finite size effects and strain rate sensitivity}. The paper is organized as follow: in section~\ref{Sim} we show the details of the molecular dynamic simulation. In section~\ref{MacroAnalysis} we present the measurements of the global thermal and mechanical response as a function of the temperature and of the strain rate. In section~\ref{MicroAnalysis}, we analyse the thermo-mechanical behaviour at the atomic scale. Finally, we identify in section~\ref{Laws} an effective thermo-mechanical constitutive law at a continuum level, able to reproduce the temperature changes when it is combined to well defined local heating sources. We finally draw conclusions in section~\ref{Concluding}.  


\section{Simulation details}\label{Sim}
\begin{figure}[h]
\centering
\includegraphics[scale=0.35]{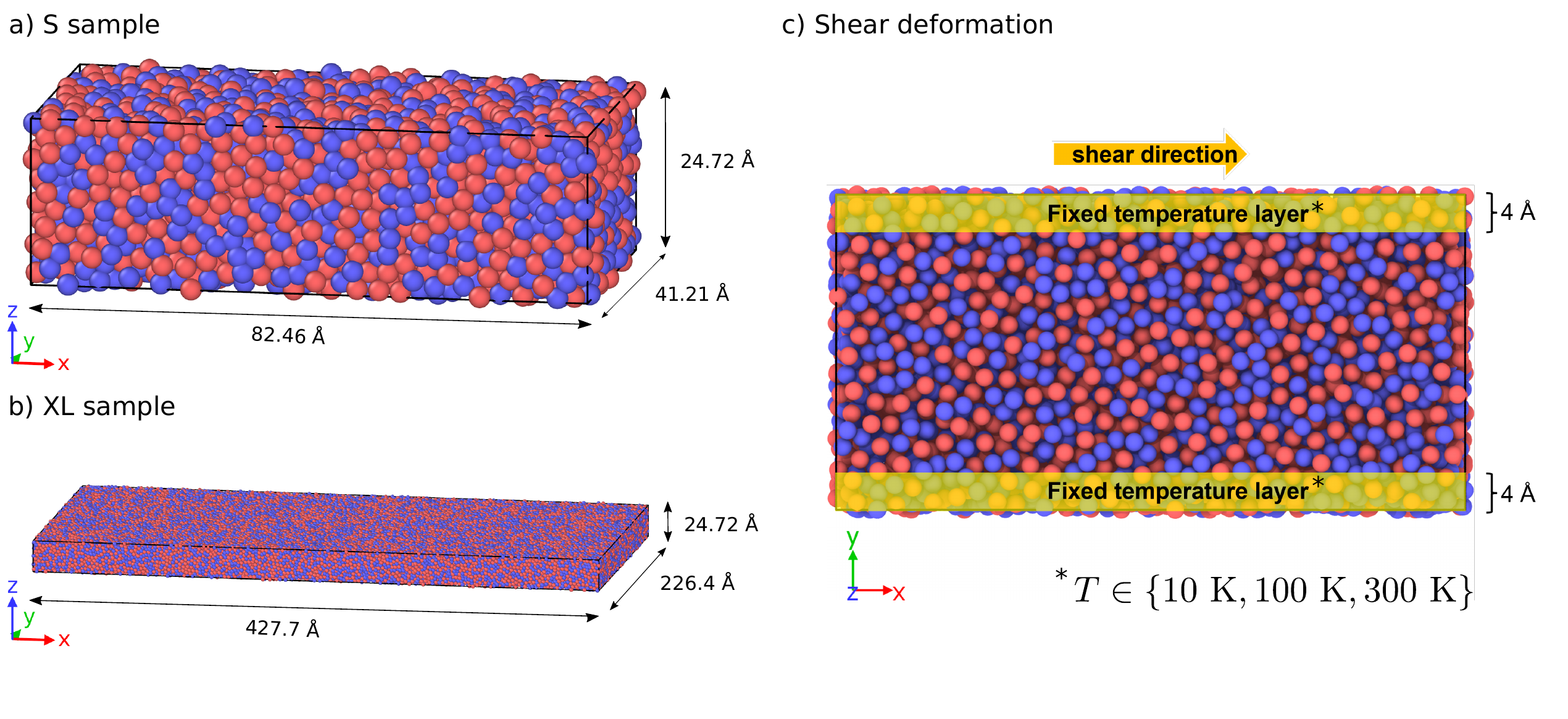}
\caption{Zr$_{50}$Cu$_{50}$ metallic glass samples considered in this work. a) Consisting of 4 800 particles, labelled as S, and b) for 145 200 atoms, labelled as XL. c) Two layers of 4~\AA~at the top and the bottom of the sample, on the $xz$ plane, are kept at constant temperature.}\label{samples}
\end{figure}

We perform molecular dynamics simulations of Zr$_{50}$Cu$_{50}$ metallic glass samples submitted to successive cycles of volume--preserving shear deformation. The model consists of two samples, one with 4 800 atoms with dimensions $L_x\times L_y\times L_z = 82\times 41\times 24$~\AA$^3$ \MS{(small sample, S)} and the second with 145 200 atoms with dimensions of $L_x\times L_y\times L_z = 427\times 226\times 24$~\AA$^3$ \MS{(large sample, XL)}, as shown in Fig.~\ref{samples}. The dimensions are chosen to prevent finite size effects in the elastic regime~\cite{Tanguy2002}. The samples are obtained from an initial FCC copper structure in which half of the atoms were replaced by zirconium. The resulting configuration is heated to 2200 K ensuring a constant 0 GPa external pressure during all the simulation. The integration timestep is set at $\Delta t = 1$~fs. Then, the molten metal is cooled down to 10 K following the protocol described by Wang {\it et al.}~\cite{Wang2012}. Finally, to obtain a well-equilibrated glass sample, the system evolves in NVE ensemble during 100 ps with a final minimization to ensure that all atomic forces were under 10$^{-4}$~eV$\cdot$\AA$^{-1}$. The resulting cooling rate is 10$^{12}$~K$\cdot$s$^{-1}$. The particles interact via the modified embedded-atom method (MEAM) potential~\cite{Baskes1992}, which consists of an extension of the original embedded-atom method with the addition of the angular forces. This procedure has already been used to explore different properties of Zr$_{50}$Cu$_{50}$ metallic glass samples, like the onset of plasticity~\cite{SepulvedaMacias2016} or plastic failure~\cite{SepulvedaMacias2020}.  All simulations were carried out using LAMMPS~\cite{LAMMPS} and the visualization using the OVITO software~\cite{Stukowski2010}.

In order to follow the temperature evolution on the sample during the cycles of shear, we mimic the experimental conditions, where external boundaries are at constant temperature by fixing two layers of atoms with a thickness of 4~\AA, one on the top of the $xz$--plane and one on the bottom of the $xz$--plane, as can be seen in Fig.\ref{samples}(c), where the shear deformation direction is also shown. \MS{The temperature is imposed through a Nos\'{e}-Hoover thermostat with damping parameter $\tau_T = 2 fs$.}
Three temperatures were considered for the fixed layers, $T_0\in\{10~\text{K}, 100~\text{K}, 300~\text{K}\}$, and for each temperature 10 statistically independent samples were generated. These samples have been obtained from the following procedure: First, we created 10 velocity profiles using a gaussian distribution with a random seed to desired temperature $T_0$. Then the samples were heated to 800 K with constant 0 GPa external pressure during 300 ps in the NPT ensemble. A NVE equilibration during 100 ps has been run at 800 K, followed by a cooling to final desired temperature ($10~\text{K}$, $100~\text{K}$, $300~\text{K}$) during 300 ps in the NPT ensemble. A final NVE equilibration for 50 ps give us the final configuration samples considered for this study, ensuring a thermodynamical equilibrium in the initial state. All those samples were then subjected to three cycles of volume--preserving shear deformation at three imposed shear rates $\dot\gamma \in\{10^8, 10^9, 10^{10}\}$~s$^{-1}$. For this, the simulation cell is deformed using a constant engineering shear strain rate, where the tilt factor changes linearly with time from its initial to final value. Volume preserving, periodic boundary conditions are applied in \MS{all three directions}. The simulation cell is deformed every 1000~$\Delta t$ for S and every 5000~$\Delta t$ for XL samples. This protocol makes it possible to ensure that mechanical waves reach the other end of the system before propagating a next wave, thus avoiding an accumulation of mechanical energy that would make our system less realistic. Finally, the fixed layers kept the constant temperature $T_0$ during all the simulation in the NVT ensemble while atoms outside the layers are controlled by NVE during all the deformation process. 


\section{Measurement of the thermal and mechanical response at the global scale}\label{MacroAnalysis}
In the following we show evidence, from the macroscopic stress--strain curves, of the role of the external temperature, the strain rate and of finite size effects on MG samples subjected to cycles of shear. For each temperature and each shear rate, we calculated the average stress--strain curve and the average temperature, for the atoms outside the fixed layers, over ten configurations. \MS{The global average temperature is given by the average of the kinetic energy fluctuations on all atoms, after the streaming velocity of the atoms caused by the change in shape of the simulation box has been removed, as well as the rigid body motion of the mobile atoms.} These results are summarized in Fig.~\ref{S_results}. \MS{The stress-strain behaviour is presented on top of each plot and on bottom we presented the temperature evolution as a function of the strain. Red solid lines of each plot represent the loading regime and green solid curves the unloading regime. The three strain rates and temperatures $T_0$ are summarized in Fig.~\ref{S_results}.}
The shear modulus $\mu$ is then measured on each case using a linear fit of the shear stress-shear strain curve over the initial 0.05 strain window. The values are summarized in Table~\ref{tbl1}.

\subsection{Small Size Samples}

\begin{figure}[h]
\centering
\includegraphics[scale=0.05]{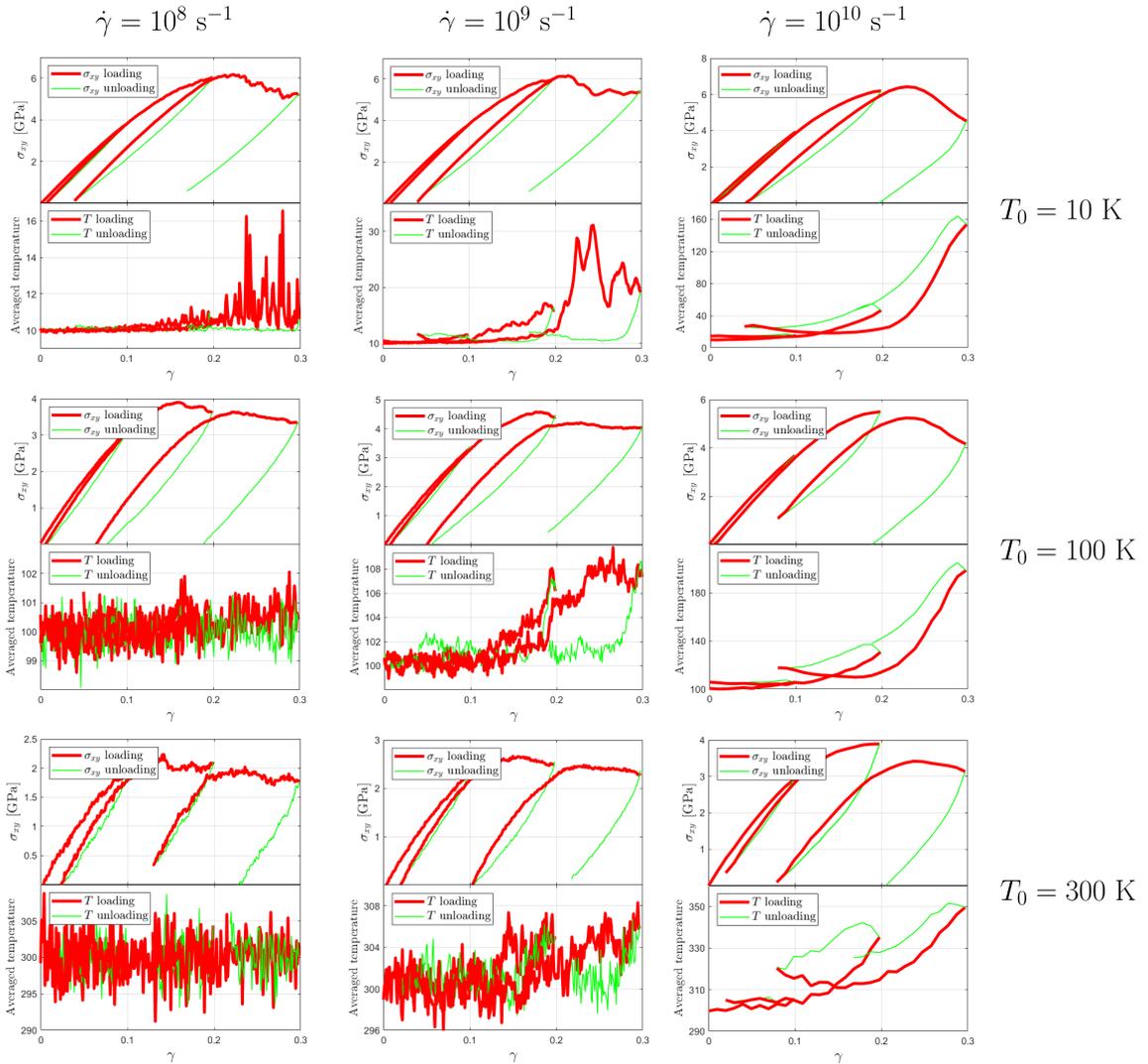}
\caption{\MS{The averaged stress-strain curves and temperature evolution of Zr$_{50}$Cu$_{50}$ metallic glass sample S with 4 800 atoms are presented. The stress and temperature results for each $T_0$ value are presented as a function of strain for the three strain rates $\dot\gamma \in\{10^{8}, 10^9, 10^{10}\}$~s$^{-1}$. The stress evolution is shown on top, while the temperature is shown on the bottom. The red lines indicate the loading regime, and the green lines indicate the unloading regime.}}\label{S_results}
\end{figure}

At lowest strain rate, $\dot\gamma = 10^8$~s$^{-1}$ (first column on Fig.~\ref{S_results}) the mechanical behaviour of the \MS{S} samples follows a quasistatic-like deformation regime. This very high value for the smallest strain rate may be surprising (apart from high power laser-induced shocks, for which it can reach $10^6$~s$^{-1}$~\cite{Raffray2023}, in up-to-date mechanical experiments, it varies more generally from $10^{-2}$~s$^{-1}$ to $10^3$~s$^{-1}$~\cite{Kermouche2022}), but such an apparently high strain rate value has already been considered representative of quasistatic deformation~\cite{Fusco2014}. This may be due to the use of empirical potentials in the simulations, probably underestimating the amplitude of energy barriers, such that the low numerical time scale corresponds to larger experimental \MS{effective} time scales. In this {\it low strain rate} case, the plastic collective rearrangements of atoms are more pronounced, as visible in the large stress drops making the stress-strain curve noisier as already shown and related to the local nature of irreversible rearrangements~\cite{Tanguy2006,Maloney2006,Fusco2010}. It is possible to identify two cases, the first, at a very low temperature ($T_0 = 10$~K, top panel of the first column on Fig.~\ref{S_results}) the system exhibits a clear increase in the temperature rises when the plastic events develop, concentrated in the region of $\gamma \in [0.20,0.30]$. The second case is related to the samples at $100~\text{K}$ and $300~\text{K}$, here, the fluctuations in temperature do not allow finding a pattern like the previous one. The increases in temperature as a function of $\gamma$ are very noisy \MS{whatever the $\gamma$ values} and not greater than $2~\text{K}$ for the $100~\text{K}$ case, while at room temperature the variations are irregular and not greater than $10~\text{K}$ during the three cycles of shear.

As we increase the strain rate to $\dot\gamma = 10^{9}$~s$^{-1}$ (second column on Fig.~\ref{S_results}) the mechanical behaviour becomes less noisy than in the previous case. The coherent stress drops shown in the coldest case at $10~\text{K}$ (see, for instance, $\gamma\approx0.22$) becomes less clear for $100~\text{K} $ and $300~\text{K}$. The shear cycles are however related in \MS{all cases} to a thermal hysteresis, with significant increases as plastic events develop throughout the sample. It shows changes in temperature ranging from $30~\text{K}$ (upper figure) to $10~\text{K}$ for the lower figure \MS{(highest $T_0$)}, becoming less evident for the latter.

Finally, at fastest strain rate, $\dot\gamma = 10^{10}$~s$^{-1}$ (third column on Fig.~\ref{S_results}) the global thermo--mechanical behaviour is similar for the three temperatures under study. In this case, the mechanical behaviour is similar, while less noisy and related to higher stress values, as in the previous cases, but the global averaged thermal behaviour exhibits very well defined loops during each loading-unloading cycle (also sketched in $\dot\gamma = 10^{9}$~s$^{-1}$). This phenomenon has been previously studied by~\cite{Chrysochoos1992} on duraluminium samples during a series of loading-unloading excitations at room temperature, both, from experimental and theoretical point of view. The temperature raise is however far larger in the current amorphous samples (reaching few tens of Kelvins in each computed case) \MS{than for smaller strain rates}.

\begin{table}[h]
\caption{Shear modulus $\mu$, in GPa, for each temperature and shear rate under study, averaged over 10 configurations for both sample sizes under study, and comparison (last column) with the experimental result given in~\cite{Fan2006}.}\label{tbl1}
\begin{tabular*}{\tblwidth}{@{}LLLLLL@{}}
\toprule
Size & Temperature & $10^{8}$~s$^{-1}$  &  $10^{9}$~s$^{-1}$ & $10^{10}$~s$^{-1}$ & Exp.~\cite{Fan2006}\\ 
\midrule
  &10 K  & 39.705 & 38.407 & 41.025 & \\
S &100 K & 34.644 & 33.412 & 40.639 & \\
  &300 K & 23.554 & 30.100 & 33.339 & 31.3\\
\midrule
   & 10 K  & 40.005 & 41.869 & 43.771 & \\
XL & 100 K & 36.883 & 39.319 & 33.118 & \\
\bottomrule
\end{tabular*}
\end{table}

\subsection{Larger Samples and Finite Size effects}

\begin{figure}[h]
\centering
\includegraphics[scale=0.05]{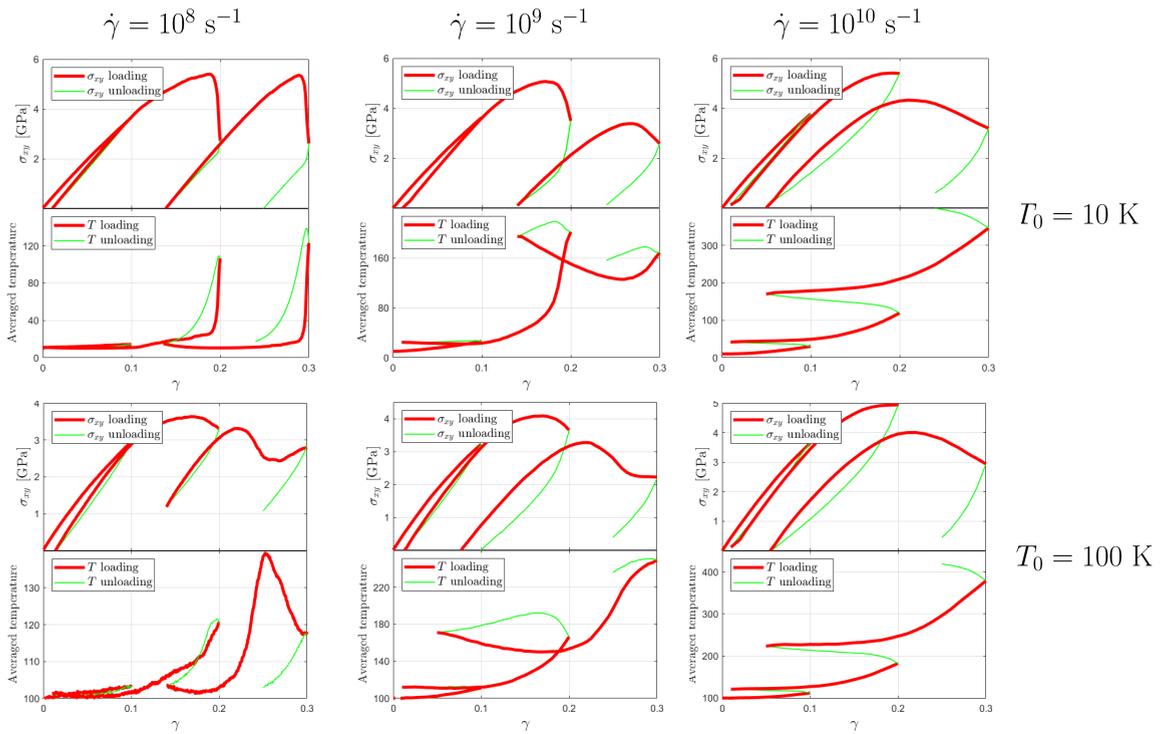}
\caption{\MS{The averaged stress-strain curves and temperature evolution of Zr$_{50}$Cu$_{50}$ metallic glass sample XL with 145 200 atoms are presented. The stress and temperature results for each $T_0$ value are presented as a function of strain for the three strain rates $\dot\gamma \in\{10^{8}, 10^9, 10^{10}\}$~s$^{-1}$. The stress evolution is shown on top, while the temperature is shown on the bottom. The red lines indicate the loading regime, and the green lines indicate the unloading regime.}}\label{XL_results}
\end{figure}

The same procedure is applied for the big sample of 145 200 particles. The stress--strain curve and the temperature evolution for the particles outside the fixed layers have been computed. The results are presented in Fig.~\ref{XL_results}, from left to right for $\dot\gamma = 10^{8}$~s$^{-1}$, $\dot\gamma = 10^{9}$~s$^{-1}$ and $\dot\gamma = 10^{10}$~s$^{-1}$ respectively.
Fig.~\ref{XL_results} shows a clear difference with respect to the small system. \MS{First,} a clear stress overshoot is seen when we apply a slow strain rate, like $\dot\gamma = 10^{8}$~s$^{-1}$. The same phenomenon, but with less intensity, is shown for $\dot\gamma = 10^{9}$~s$^{-1}$, while for the fastest deformation rate the stress softening is progressive as for the smaller system, related as will be s\MS{h}own later to an accumulation of plastic events that do not manage to form a mature shear band. More importantly, a critical point in this case is the absence of the temperature hysteresis that was shown for the small system in Fig.~\ref{S_results}. \MS{This is related to a difficulty to evacuate heat in the large systems, as will be discussed and solved later.} For the \MS{smaller strain rates in the} large system, there is a drastic change in temperature correlated with the large stress drops, as clearly seen in the lowest strain rate case. In this case, after the temperature increase, during the unloading process, the system can return to the initial system temperature, despite an increase in temperature of more than $100~\text{K}$ . For $\dot\gamma = 10^{9}$~s$^{-1}$ case, although there is a significant increase in temperature, beyond $200~\text{K}$, the system fails to release the temperature throughout the sample. At the highest strain rate, this effect is even more pronounced: in this latter case, while the decrease in the stress as a function of strain is softer than the case $\dot\gamma = 10^{8}$~s$^{-1}$, the temperature increases monotonously \MS{with} the deformation as a function of time, even reaching up to $400~\text{K}$ at $30\%$ strain. This situation represents a substantive difference with respect to small samples and looks \MS{clearly} unphysical.

\begin{figure}
\includegraphics[scale=0.31]{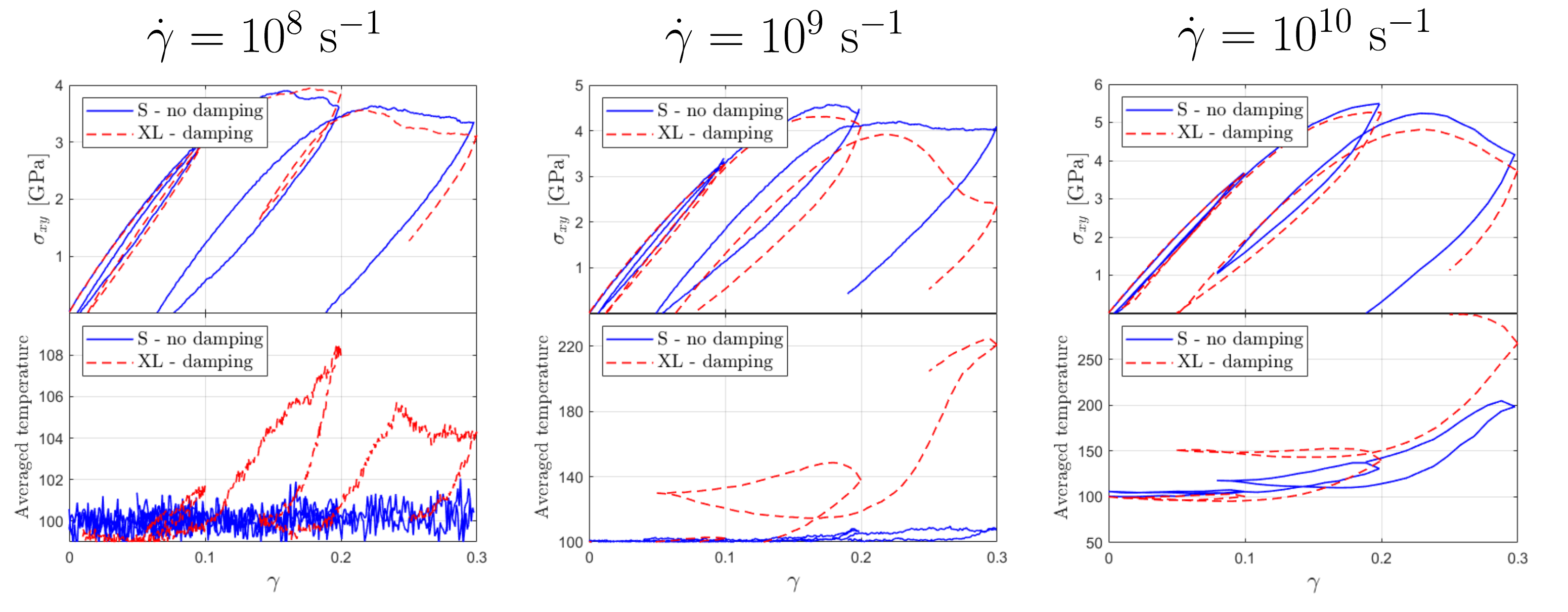}
\caption{Thermo-mechanical evolution for the big XL sample with additional uniform damping vs. small sample without damping, at $T_0 = 100~\text{K}$. From left to right $\dot\gamma \in\{10^8, 10^9, 10^{10}\}$~s$^{-1}$. From left to right the values of the damping coefficient are: $\eta=0.000001~\text{eV}\cdot\text{ps}/\text{\AA}^2$, $\eta=0.00001~\text{eV}\cdot\text{ps}/\text{\AA}^2$ and $\eta=0.0001~\text{eV}\cdot\text{ps}/\text{\AA}^2$.}\label{Damping_SvsXL}
\end{figure}

To solve this problem, it is needed to allow the system to release the energy across its size, as we apply the shear strain for the fastest strain rate case. \MS{In plus of giving sufficient time between successive strain steps to let the boundaries evacuate energy, additional quantum dissipation sources are considered.} For that, we applied a global damping force $F_i = -\eta\cdot v_i$ on each $i$-atom, where $\eta$ is the damping coefficient and $v_i$ the velocity of each particle. This additional damping force has a physical meaning: it corresponds to the physical damping induced \MS{for example} by the Fermi rule when electronic excitations/deexcitations\MS{, resulting from bond breaking,} are considered~\cite{Review1983}. As shown in Fig.~\ref{Damping_SvsXL}, it is possible this way for the largest strain rate, to get results \MS{very similar} to those obtained with the small system. This procedure is applied to the three strain rates and two temperatures presented in Fig.~\ref{XL_results}, where, for each case, the value of the damping coefficient that best fits the results of the small system shown in Fig.~\ref{S_results} is chosen independently. 
\begin{figure}
\includegraphics[scale=0.05]{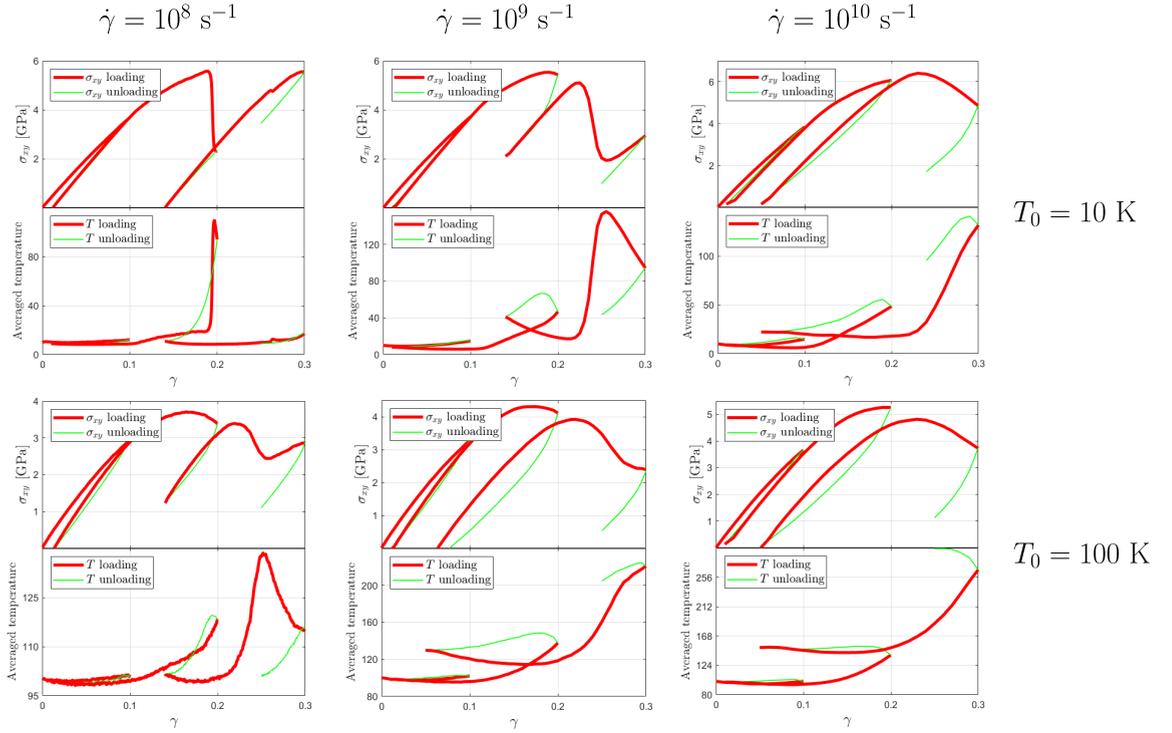}
	\caption{Thermo-mechanical evolution for the big XL sample with the introduction of a volumic viscous damping. Results for $T_0 = 10~\text{K}$ on the top panel, and for $T_0 = 100~\text{K}$ on the bottom, for the three strain rates under study. From left to right $\dot\gamma \in\{10^8, 10^9, 10^{10}\}$~s$^{-1}$. From left to right the values of the damping coefficient are, for top layer: $\eta=0.00001~\text{eV}\cdot\text{ps}/\text{\AA}^2$, $\eta=0.0001~\text{eV}\cdot\text{ps}/\text{\AA}^2$ and $\eta=0.001~\text{eV}\cdot\text{ps}/\text{\AA}^2$. For bottom layer: $\eta=0.000001~\text{eV}\cdot\text{ps}/\text{\AA}^2$, $\eta=0.00001~\text{eV}\cdot\text{ps}/\text{\AA}^2$ and $\eta=0.0001~\text{eV}\cdot\text{ps}/\text{\AA}^2$.}\label{DampingXL}
\end{figure}
The results \MS{obtained with this additional damping} are displayed in figure~\ref{Damping_SvsXL}. Here, solid lines, both red and blue, represent the loading regimes, while the dashed lines show the unloading regime. An important difference is visible in the thermal and mechanical behaviour of the S versus XL samples when they are subjected to the lower shear rates of $\dot\gamma=10^8$~s$^{-1}$ and $\dot\gamma=10^9$~s$^{-1}$. In this last case for example, the small system shows a plastic regime dominated by successive stress drops, while the large system shows a drastic single but large stress drop, usually related to the formation of a shear band. The increase in the temperature is different as well in both cases: the big sample exhibits a huge increase in the temperature correlated with the stress drops, while the small samples show small increases in temperature that are rapidly damped. Finally, for $\dot\gamma=10^{10}$~s$^{-1}$ there is good accordance between the self-heating in the small samples, and the self-heating in the large sample. This means that, for a reason that we will try to explain later, there are strong finite size effects in the plastic response of metallic glasses, \MS{especially for low strain rates. Moreover,} it is important to have a good knowledge of the dissipation sources\MS{, including quantum contributions that could be taken into account through effective parameters,} before performing classical simulations. The whole set of curves obtained for the large sample, at two different temperatures, with additional homogeneous damping is shown in Fig.~\ref{DampingXL}. \MS{This set of data will be used for the thermo-mechanical analysis performed in the next part.} 

\begin{figure}[h]
\centering
\includegraphics[scale=0.5]{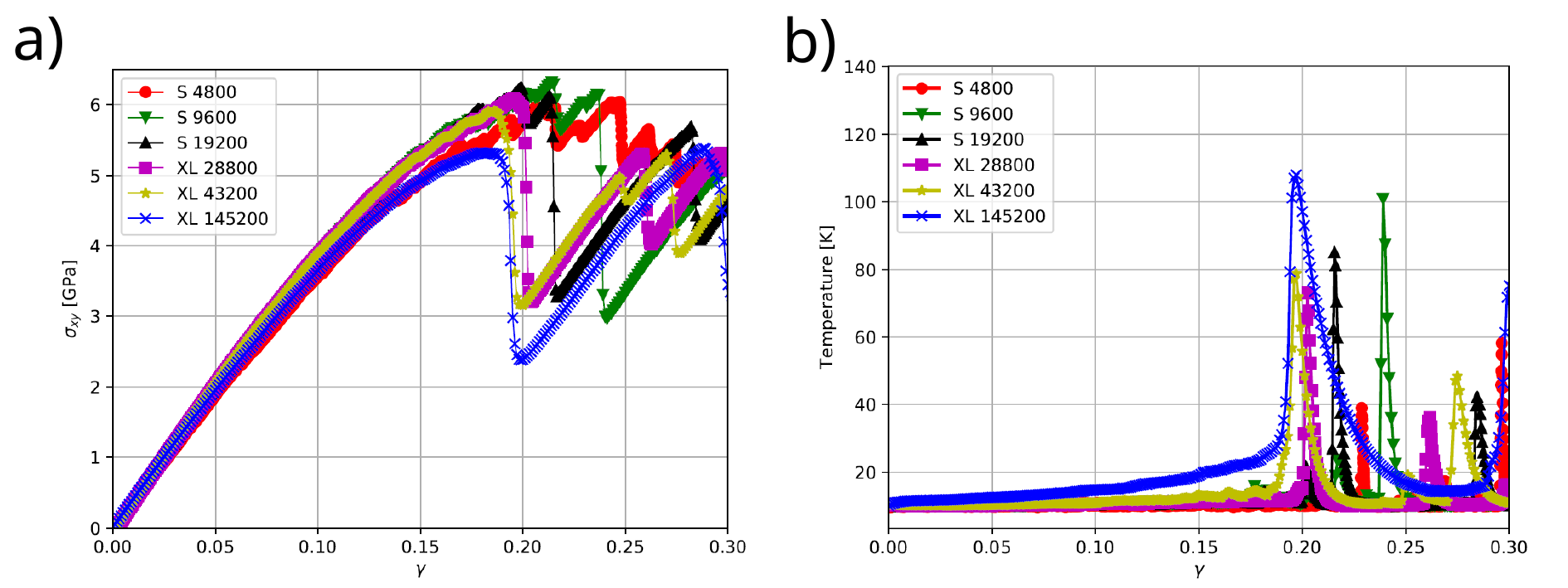}
\caption{Study of the size effect on the global thermo-mechanical behaviour of Zr$_{50}$Cu$_{50}$ metallic glass at initial 10~\text{K} and $\dot\gamma= 10^{10}$~s$^{-1}$. a) Shows the stress-strain curve for different sizes samples (number of atoms in the legend) and b) the temperature evolutions for the atoms outside the fixed layers.}\label{SizeEffect}
\end{figure}

In order to look for a macroscopic behaviour \MS{and finite size effects}, we have increased progressively the size of the system and looked at the thermo-mechanical response for the different sizes. They are superposed for six different samples sizes in the Fig.~\ref{SizeEffect}. For each case we replicated the initial S sample \MS{with the aim of getting samples with increasing sizes,} and we created a new Gaussian velocity distribution at 10~K. Then the system is heated up to 800~K at constant 0 GPa pressure during 300 ps, followed by an NVE equilibration during 100 ps and finally it is cooled down during 300 ps to reach the target temperature of 10~K. A final equilibration in NVE for 50 ps is done to ensure a well-equilibrated \MS{homogeneous} sample. \MS{This very low temperature is chosen here as a preliminary test, in order to avoid unphysical uncontrolled self-heating as discussed in the previous part for large samples.} Each of these configurations is \MS{then} subjected to a shear deformation at a strain rate of $\dot\gamma = 10^{10}$~s$^{-1}$ under the same conditions as for the S sample described in section~\ref{Sim}, except that a single run is performed without loading-unloading cycles.
Fig.~\ref{SizeEffect}(a) shows the mechanical behaviour while Fig.~\ref{SizeEffect}(b) shows the evolution of the temperature during the deformation process, for each sample size. The related number of atoms \MS{in the sample} is visible in the legend of each plot. The Fig.~\ref{SizeEffect}(a) shows clearly a transition from a noisy behaviour for sample S (red curve), with multiple small stress drops after $\gamma=0.15$, to a continuous stress softening with a large stress \MS{drop}, as shown for the XL sample (blue curve). This transition is accompanied by a progressive increase in the amplitude of the stress drops starting from the plastic flow regime. A larger size seems necessary to allow the setting up of the complete stress relaxation, that occurs finally at slightly smaller global strain when the system size increases. This means that our smallest sample S is not an elementary representative volume for the plastic deformation. In terms of temperature, Fig.~\ref{SizeEffect}(b) shows a difference $\Delta T \approx 50$~K in temperature change between samples S and XL. While sample S exhibits a succession of small, irregular temperature increases, related to the occurrence of small plastic events throughout the simulation, XL sample shows a significant smooth temperature increase, marked by large amplitude stress softening, and guided by self-heating. \MS{The thermal behaviour becomes smoother with the system size, as in self-averaged processes, but it does not clearly converge to a limit behaviour, even for the largest sample, suggesting strong finite size effects that are difficult to catch with Molecular Dynamics simulations. In the following, we will thus keep the two sizes related to the S and XL samples, in all our analysis, for comparison.}

\section{Analysis of thermal behaviour at the micro-scale}\label{MicroAnalysis}

Once the global thermo-mechanical behaviour has been measured, we now focus on the visualization at the atomic scale of the thermal response and the strain localization on both, S and XL, samples. For this, we calculate the local infinitesimal small shear strain using the per-particle coarse graining (CG) strain tensor~\cite{Goldhirsch2002,Goldenberg2007,Goldenberg2008,Tsamados2009}, with a coarse graining width of $\omega_{\text{cg}} = 6$~\AA. \MS{This value is chosen with the need to avoid displacement and strain singularities in the coarse-grained quantities. As shown in Ref.\cite{Tsamados2009, Molnar2016}, it corresponds to at least two times the first neighbours distance, close to the size of the core of the elementary plastic events in these systems~\cite{Albaret2016}.} The CG strain tensor is obtained as
\begin{equation}\label{cgstrain}
\varepsilon_{\alpha\beta}({\bf r},t) = \frac12\left(\frac{\partial u_{\alpha}({\bf r},t)}{\partial r_{\beta}} + \frac{\partial u_{\beta}({\bf r},t)}{\partial r_{\alpha}}  \right) 
\end{equation}
where the {\it coarse-grained} continuous displacement field is obtained from
\begin{equation}
{\bf u}({\bf r},t) = \frac{\sum_i m_i{\bf u}_i(t)\phi[{\bf r} - {\bf r}_i(t)]}{\sum_j m_j\phi[{\bf r} - {\bf r}_j(t)]}
\end{equation}
Here, $\phi[{\bf r}]$ is the coarse-grained function which corresponds to a normalized Gaussian function of width $\omega_{\text{cg}}$\MS{, $u_i(t)$ is the displacement of atom $i$ at time $t$, and $m_i$ is its mass}. The time interval used to compute the strain from the displacement is 1 ps, this corresponds to a strain interval between initial and final configuration of $\Delta \gamma = 0.01$ for  $\dot\gamma = 10^{10}$~s$^{-1}$, and more generally $\Delta\gamma = 10^{-12}\dot\gamma$ when the unit used for $\dot\gamma$ is s$^{-1}$.
For the sake of simplicity and because it is the main signature of plasticity in metallic glasses~\cite{Tanguy2021}, we visualize here only the deviatoric part of the strain tensor located on each atom (${\bf r} = {\bf r}_i $). For this we diagonalize the CG strain matrix and consider the maximum absolute value of the difference between the three eigenvalues of $\varepsilon$ tensor. Also, the local temperature is computed, on each atom $i$, from its kinetic energy $K_E (i) = 1/2 m_i v(i)^2 = 3/2 k_B T (i)$ computed after having subtracted the global shear flow, $m_i$ being the mass of the atom, $v (i)$ its velocity, $k_B$ the Boltzmann constant, and $T (i)$ the resulting temperature on atom $i$. 

%
\begin{figure}[h]
\centering
\includegraphics[scale=0.18]{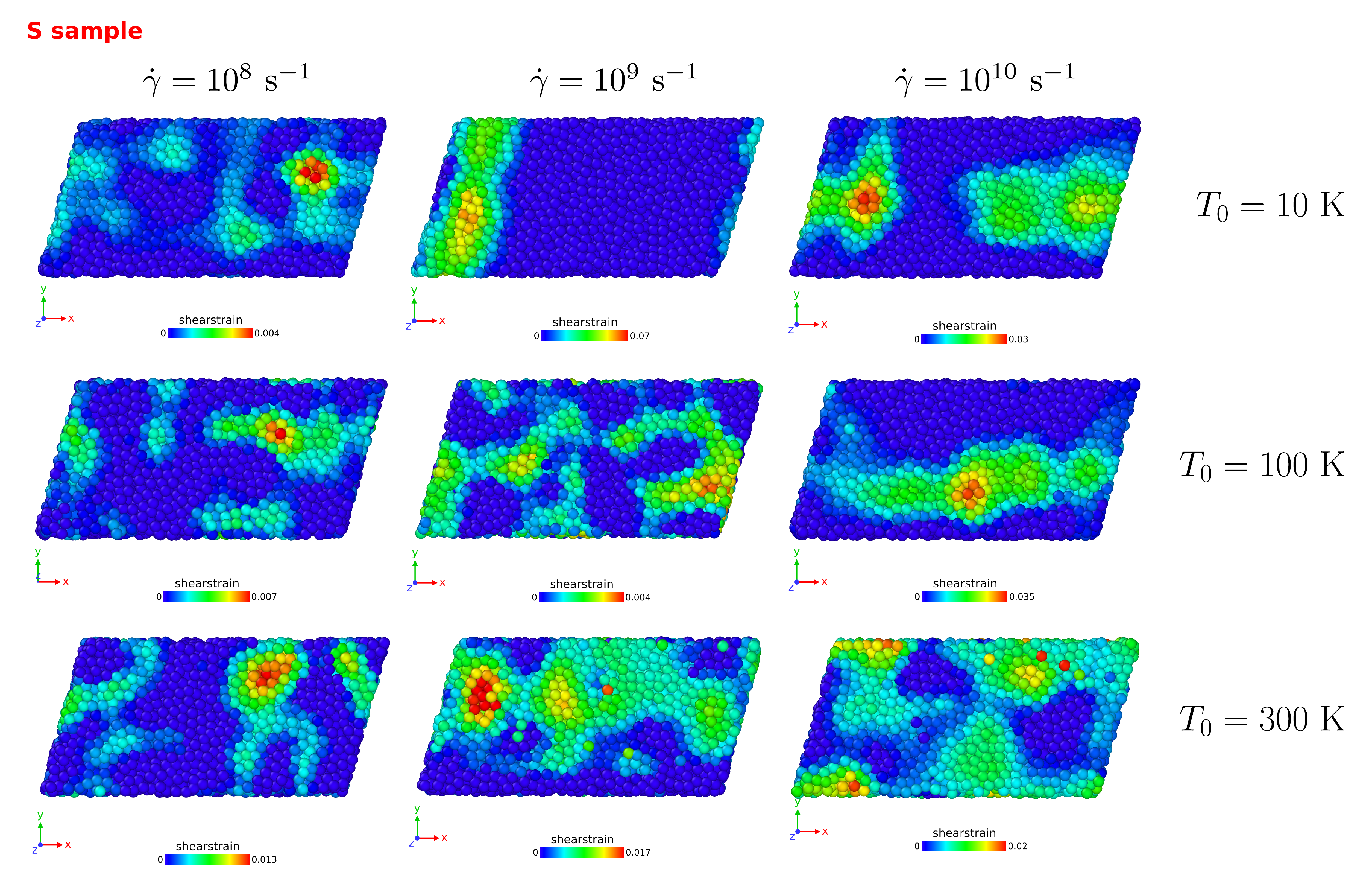}
\caption{Instantaneous shear strain localization for S sample at $\gamma=0.28$. From left to right $\dot\gamma= 10^{8}$~s$^{-1}$, $\dot\gamma = 10^9$~s$^{-1}$ and $\dot\gamma = 10^{10}$~s$^{-1}$. From top to bottom for different temperatures $10~\text{K}$, $100~\text{K}$ and $300~\text{K}$.}\label{ShearstrainS}
\end{figure}
\begin{figure}[h]
\centering
\includegraphics[scale=0.18]{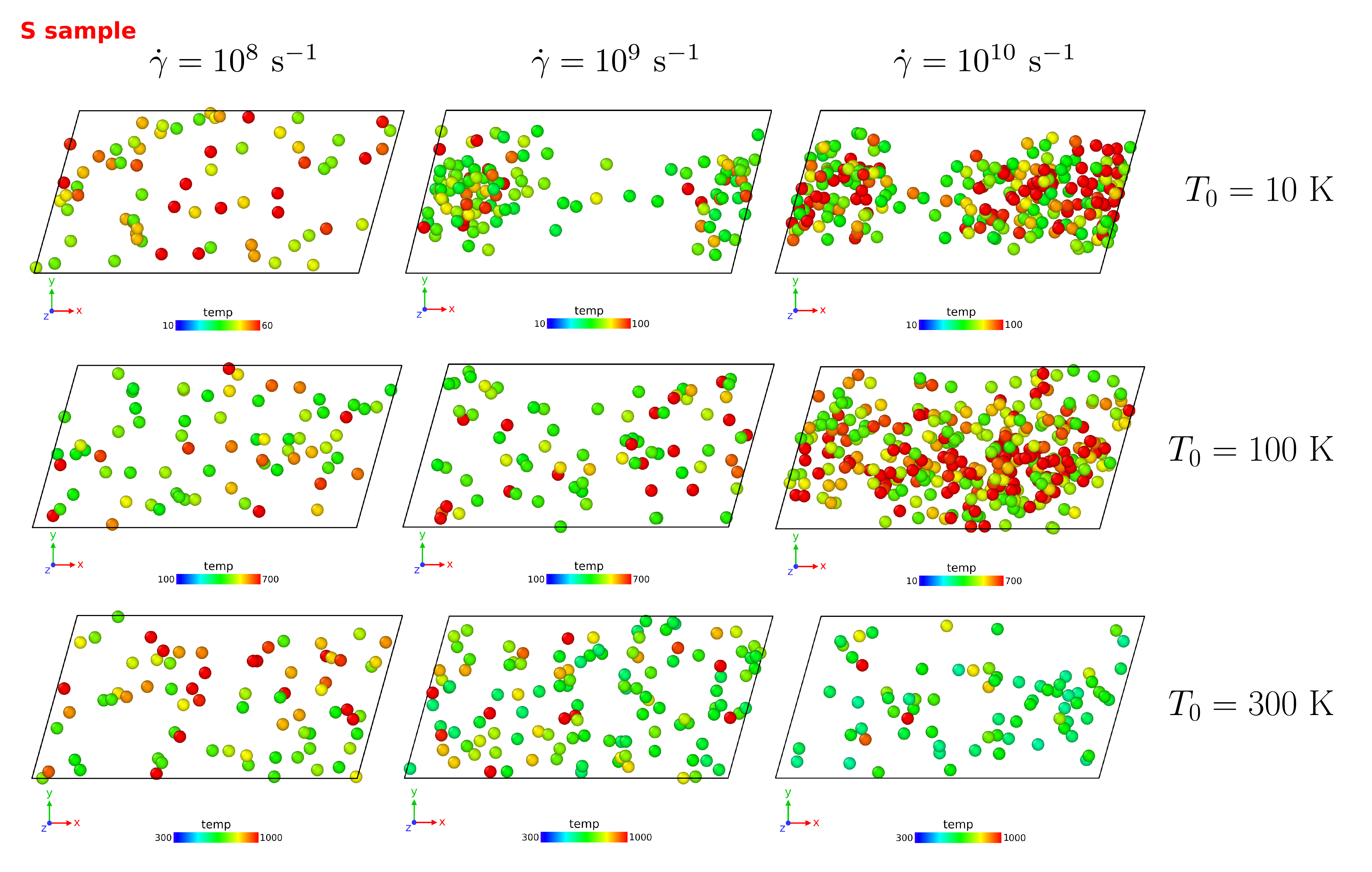}
\caption{Local temperature for S sample at $\gamma=0.28$. From left to right $\dot\gamma= 10^{8}$~s$^{-1}$, $\dot\gamma = 10^9$~s$^{-1}$ and $\dot\gamma = 10^{10}$~s$^{-1}$. \MS{For the first row ($T_0 = 10~\text{K}$) only atoms with a temperature greater than $T \geq 60~\text{K}$ are displayed. For the second row ($T_0 = 100~\text{K}$) atoms greater than $T \geq 200~\text{K}$ are displayed and finally for the last row ($T_0 = 300~\text{K}$) atoms greater than $T \geq 500~\text{K}$ are shown.}}\label{TLocalS}
\end{figure}
We show in Fig.~\ref{ShearstrainS}, the local instantaneous deviatoric strain computed from the CG strain tensor at a global strain of $\gamma=0.28$ \MS{in the S sample}, considering as the reference configuration the one that precedes the current frame in the simulation sequence. The plastic behaviour is dominated by the appearance and growth of shear transformation zones (STZs), which appear as highly strained sub-nanometric spots throughout the sample, without forming a shear band. From Fig.~\ref{ShearstrainS} only \MS{two of the nine samples ($\dot\gamma = 10^9~\text{s}^{-1}$ at $10~\text{K}$ and $\dot\gamma = 10^{10}~\text{s}^{-1}$ at $100~\text{K}$ ) exhibit} clearly a shear band at this stage of deformation\MS{, a third one ($\dot\gamma = 10^{10}~\text{s}^{-1}$ at $10~\text{K}$ ) exhibits a slightly incomplete shear band. }
Fig.~\ref{TLocalS} shows the local temperature of the samples shown in Fig.~\ref{ShearstrainS}, at the same instant of deformation. The temperature is distributed inside the sample. The three previous cases are of interest here\MS{:} at $10~\text{K}$, both $\dot\gamma = 10^9~\text{s}^{-1}$ and $10^{10}~\text{s}^{-1}$ have a higher temperature concentration in the same spots where the STZs are evolving (shear bands). This phenomenon is repeated also for the $100~\text{K}$ case with the fastest strain rate. \MS{This correspondence between local strain and temperature holds very well when the strain is localized along shear bands. For the smaller strain rates at these two temperatures, and for all the strain rates at the highest studied temperature $T=300~\text{K}$, the deformation is noisy, the STZs do not look spatially correlated in the S sample.}

\begin{figure}[h]
\centering
\includegraphics[scale=0.20]{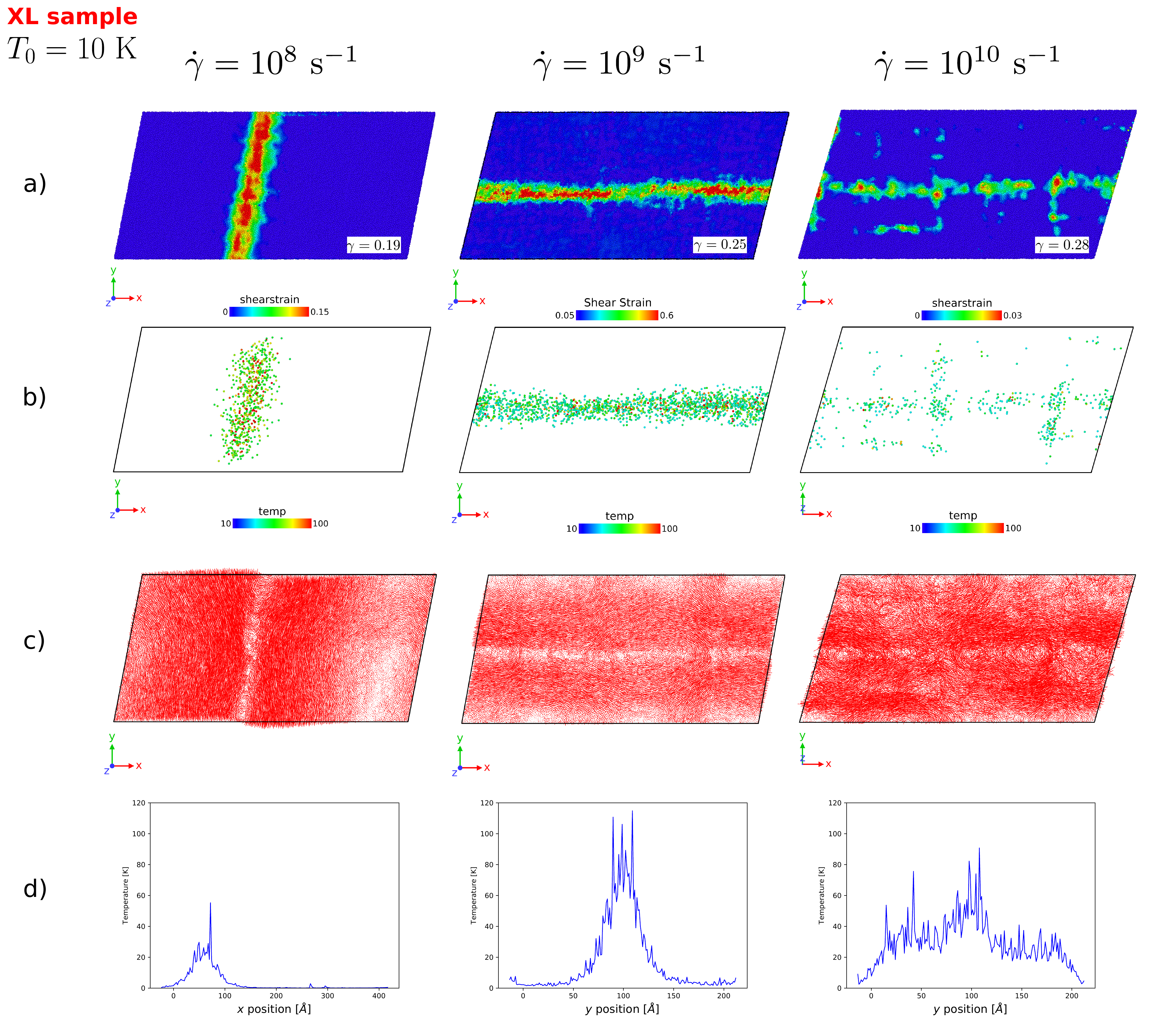}
\caption{a) Shear bands for XL sample at $T_0 = 10~\text{K}$. From left to right $\dot\gamma= 10^{8}$~s$^{-1}$, $\dot\gamma = 10^9$~s$^{-1}$ and $\dot\gamma = 10^{10}$~s$^{-1}$. Shear strain is computed from the CG deviatoric part of the strain tensor, computed from Eq.~\ref{cgstrain}. b) Snapshot of the local temperature for XL sample at the stage of formation of the shear band (SB)\MS{, here atoms with a temperature greater than $T\geq 15~\text{K}$ are displayed}. c) Non--affine displacement (NAD) field, on a slab of $4$~\AA~of thickness, on the center of the XL samples in the $z$-plane. The NAD showed corresponds to the instantaneous value computed from two successive steps ($\Delta t = 1 ps $) at $T_0 = 10~\text{K}$. d) Temperature profile as a function of the orthogonal direction of formation of the SB. The absolute maximum value for temperature is displayed}\label{SBNADTXL10}
\end{figure}
\begin{figure}[h]
\centering
\includegraphics[scale=0.20]{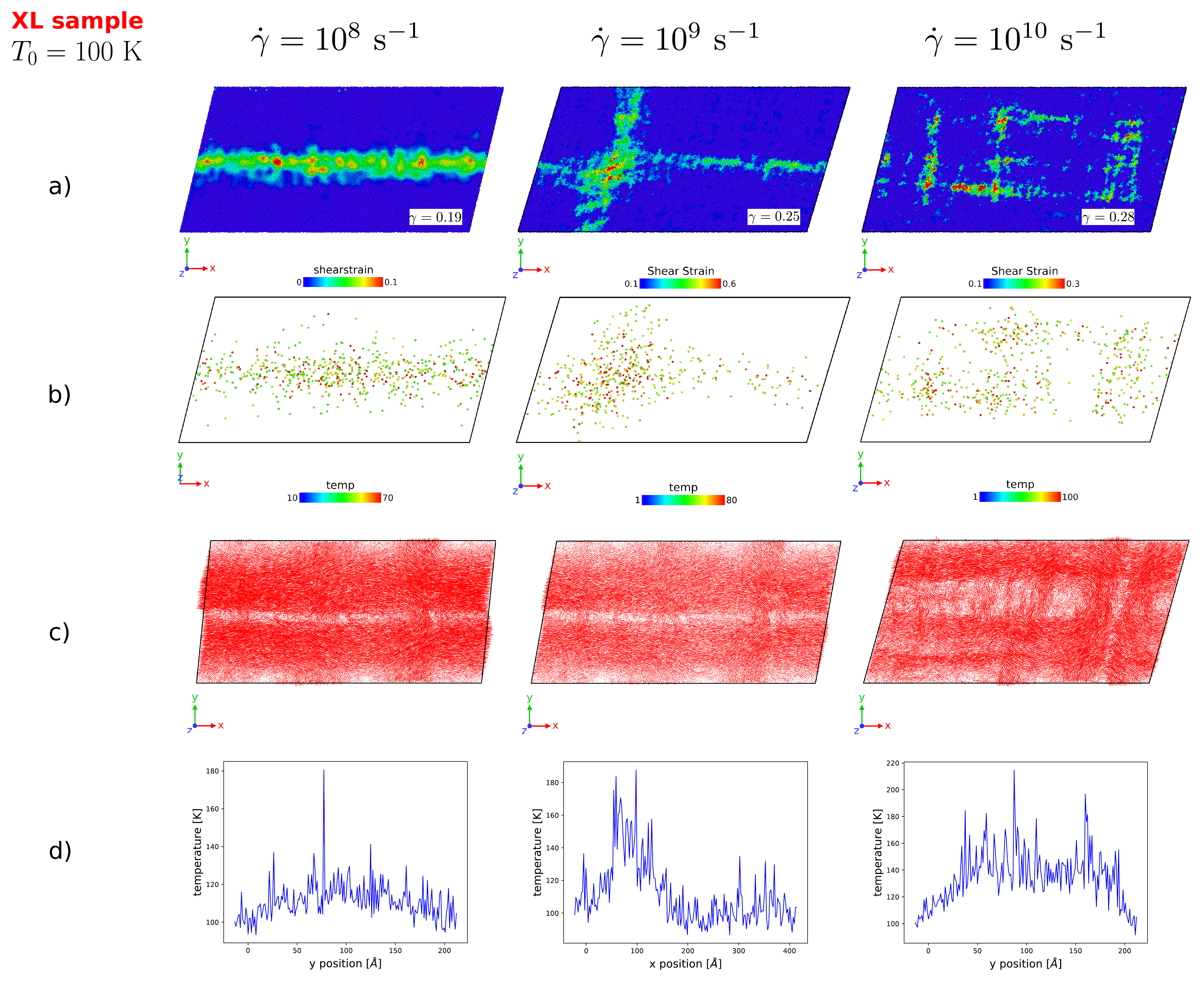}
\caption{a) Shear bands for XL sample at $T_0 = 100~\text{K}$. From left to right $\dot\gamma= 10^{8}$~s$^{-1}$, $\dot\gamma = 10^9$~s$^{-1}$ and $\dot\gamma = 10^{10}$~s$^{-1}$. Shear strain is computed from the CG deviatoric part of the strain tensor, computed from Eq.~\ref{cgstrain}. b) Snapshot of the local temperature for XL sample at the stage of formation of the shear band (SB)\MS{, here atoms with a temperature greater than $T\geq 120~\text{K}$ are displayed}. c) Non--affine displacement (NAD) field, on a slab of $4$~\AA~of thickness, on the center of the XL samples in the $z$-plane. The NAD showed corresponds to the instantaneous value computed from two successive steps ($\Delta t = 1 ps $) at $T_0 = 100~\text{K}$. d) Temperature profile as a function of the orthogonal direction of formation of the SB. The absolute maximum value for temperature is displayed.}\label{SBNADTXL100}
\end{figure}
The results for the XL sample, with the uniform damping force, for the three different strain rates already applied on the S sample, are presented in Fig.~\ref{SBNADTXL10} for $T_0 = 10~\text{K}$ and in Fig.~\ref{SBNADTXL100} for $T_0 = 100~\text{K}$. As before, for each shear rate and external temperature, we computed the instantaneous deviatoric strain. As can be seen from Fig.~\ref{SBNADTXL10}(a) at $T_0 = 10~\text{K}$, there is a strong shear band formation for both, $\dot\gamma= 10^{8}$~s$^{-1}$ and $\dot\gamma = 10^9$~s$^{-1}$ shear rates, while the highest strain rate does not present a well-defined shear band, but rather an accumulation of plastic activity distributed throughout the sample and displaying various alignments with a thinner width, as previously described in~\cite{Tanguy2002,SepulvedaMacias2018}. The same situation holds for $T_0 = 100~\text{K}$ (Fig.~\ref{SBNADTXL100}(a)), but the transition to shear banding occurs at a smaller strain rate, since the shear band is already not very well defined at $\dot\gamma = 10^9$~s$^{-1}$. This means that the mechanical behaviour of the S sample is far from having converged to the macroscopic behaviour, and that increasing the strain rate, or the temperature appears unfavourable to the formation of a global shear band, as already mentioned in the review article~\cite{Tanguy2022}. As for the S sample before, the local temperature is directly correlated to the local strain, displayed in figures~\ref{SBNADTXL10}\MS{(b)} and~\ref{SBNADTXL100}(b), suggesting that the inhomogeneous deformation is induced by an instability giving rise to local accelerations. \MS{In the XL sample, unlike in S sample, the correspondence between local deformation and temperature is valid for all the strain rates, down to the slowest strain rate studied here. At this slow strain rate, a large but well-defined shear band is already visible, unlike in S samples. The latter are clearly too small to reproduce this occurrence of these large shear bands at slow strain rates, since the width of the related shear band reaches $50$~\AA, larger than the width of the S samples. This underlines the role of the finite size in the difficulty of reproducing and identifying large scale  strain localization in the low strain rate regime. In the large samples, the} local temperature increase can reach few tens of degrees on nanometer length scales, and \MS{it is} amplified with the global strain rate. In the absence of a well defined shear band, that is for the largest strain rates \MS{only}, it is seen that the temperature increase is distributed over the whole sample, yielding a higher global self-heating. In addition, figures~\ref{SBNADTXL10}\MS{(c)} and~\ref{SBNADTXL100}(c) show the behaviour of the instantaneous non-affine displacement~\cite{Tanguy2002} at the instant of formation of the shear band. It displays shear glidings along shear bands, surrounded by \MS{mesoscale} rotational displacements, as previously shown in the literature~\cite{Tanguy2002,Sopu2017,SepulvedaMacias2020}. Finally figures~\ref{SBNADTXL10}\MS{(d)} and~\ref{SBNADTXL100}(d) show the temperature profile as a function of the orthogonal direction compared to that of the SB. The absolute maximum value for temperature is displayed in the figure. From here it is possible to extract that there is a high concentration of temperature \MS{increase} within the shear band, even evident during its formation process. There is a self-heating process of the sample as the shear band is formed, with a significant increase compared to the rest of the sample, as can be seen for example in the case of Fig.~\ref{SBNADTXL10}(d) for a shear rate of $\dot\gamma = 10^9$~s$^{-1}$. This phenomenon is in good agreement with respect to recent results presented by Lagogianni~{\it et al.}~\cite{Lagogianni2022} for the case of binary Lennard-Jones mixture. The spatial resolution of the temperature increase allows clearly identifying the width of the shear band, which is of the order of $100$~\AA, significantly larger than the size of the S sample. This confirms \MS{again} the fact, that the S sample is \MS{too} small to properly account for plastic deformation in the metallic glass samples. This could also be the case for the XL sample at large strain rates (larger shear bands are suggested for the highest strain rates in figures~\ref{SBNADTXL10} and~\ref{SBNADTXL100}(d)).

We will now relate the global self-heating of the samples to heat sources related either to the energy dissipated in the local plastic deformations, either to the internal thermo-mechanical couplings as described in \MS{a continuous parametric description of the} free energy\MS{, that will be discussed in the next part}. The relative weight of each contribution depends on the thermal and mechanical excitations, that is on the thermo-mechanical load imposed at the boundaries\MS{, especially the external temperature and the imposed strain rate}. 

\begin{figure}[h!]
\centering
\includegraphics[scale=0.45]{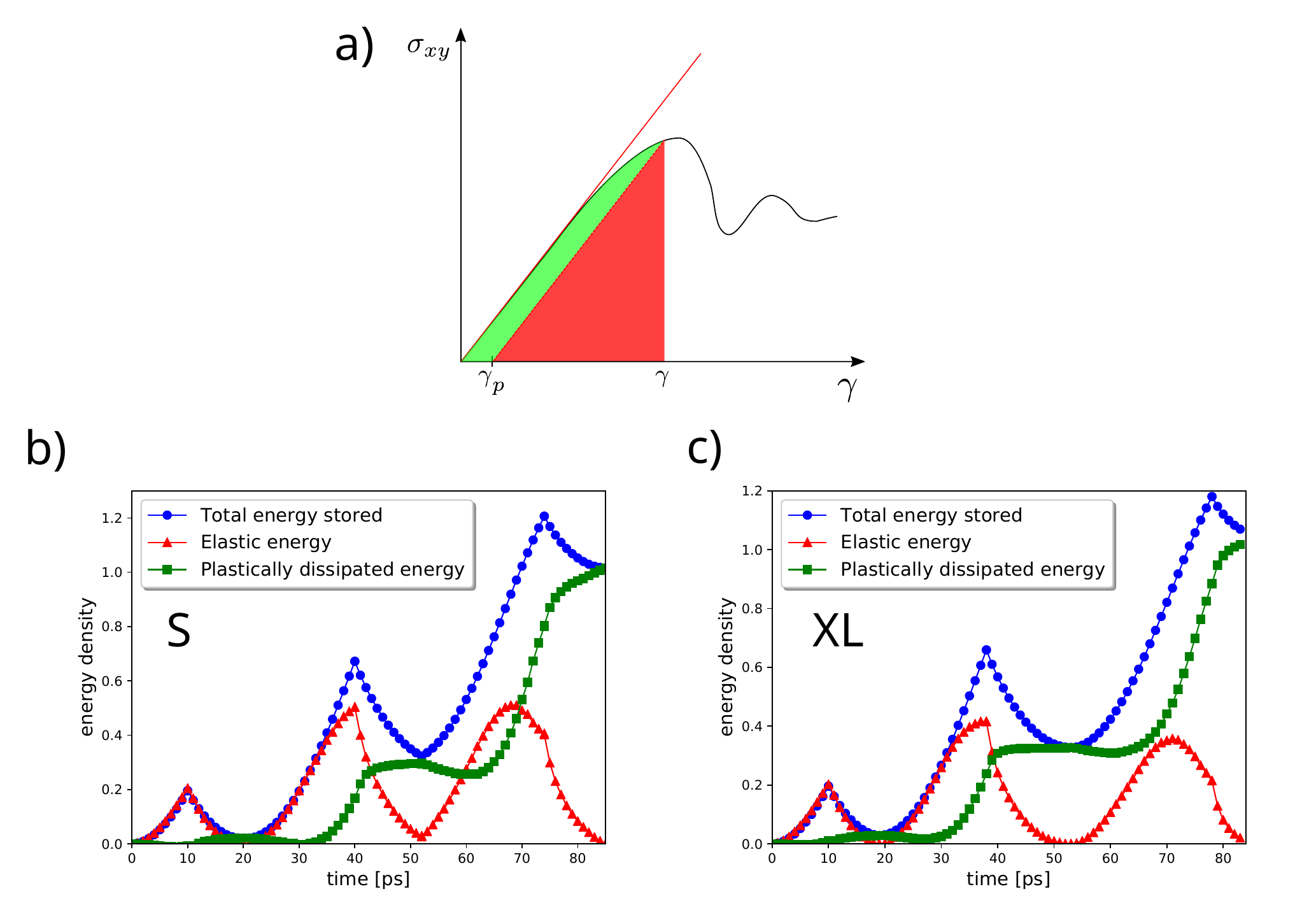}
\caption{a) Comparison between the elastic strain energy and the dissipated energy \MS{per unit volume}. The shear modulus values are given by the initial slope in the stress-strain curve and summarized in Tab.~\ref{tbl1}. The three different kinds of \MS{energy densities}: elastic, plastically dissipated, and total energy \MS{per unit volume} are computed here for $T_0 = 100~\text{K}$ and for a strain rate of $\dot\gamma=10^{10}$~s$^{-1}$. In b) result obtained for the small S sample, and in c) for the big XL sample.}\label{Scheme_Energies_S_XL}
\end{figure}
\begin{figure}[h]
\centering
\includegraphics[scale=0.3]{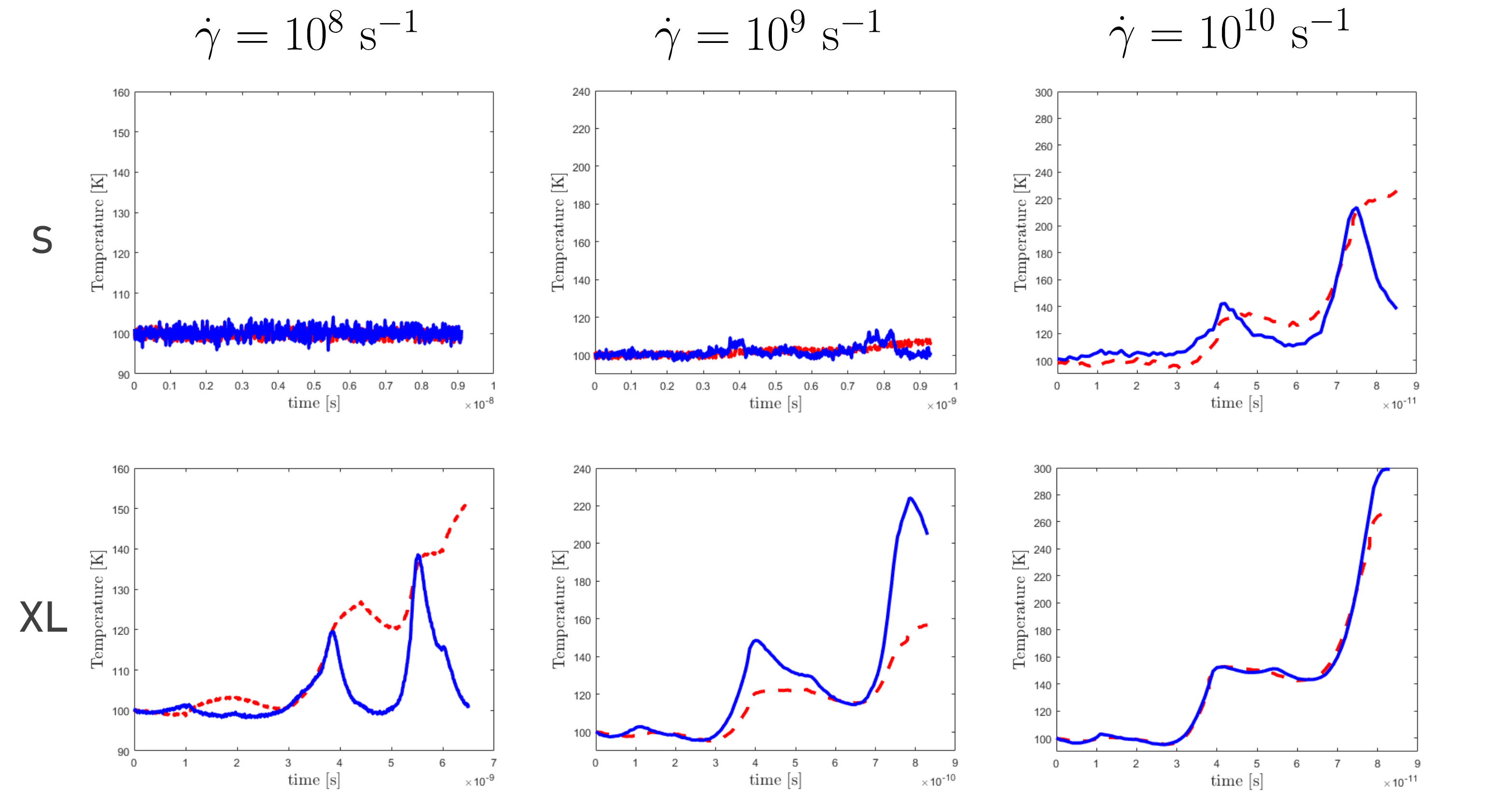}
\caption{Analysis at $T_0 = 100~\text{K}$ for S (top) and XL (bottom) samples \MS{with additional damping}. From left to right $\dot\gamma = 10^8$~s$^{-1}$, $\dot\gamma = 10^9$~s$^{-1}$ and $\dot\gamma = 10^{10}$~s$^{-1}$ respectively. The blue curve is the global temperature measured as a function of time, and the red curve is the temperature computed from the intrinsic and isentropic contributions (with local plasticity acting as a heat source and thermo-mechanical couplings helping storing heat) - as discussed in the text. The parameters used in the model are summarized in table~\ref{tbl2}.}\label{TdeT}
\end{figure}

\section{Fitting constitutive laws and thermo-mechanical couplings}\label{Laws}

In the following, we will propose a \MS{continuous} model \MS{inspired from~\cite{Chrysochoos1992}} to describe the self-heating phenomenon in connection to the energy dissipated by means of plastic deformation. At the continuum scale, we will consider here the glass as an homogeneous, isotropic, linear-thermoelastic solid. At finite temperature, the simplest expression for the strain--energy density is~\cite{Book:Sadd} 
\begin{equation}
\psi = \frac12 \sigma_{ij}\epsilon_{ij}^{\;\sigma}
\end{equation}
with $\sigma_{ij}$ the stress tensor components and $\epsilon_{ij}^{\;\sigma}$ the temperature independent -- or elastic -- contribution to the strain. Using $\sigma_{ij} = C_{ijkl}\epsilon_{kl}^{\; \sigma}$ and $\epsilon = \epsilon^{\sigma} + \epsilon^{T}$, where $C_{ijkl}$ are the elastic moduli, $\epsilon^T = \alpha_{ij}(T-T_0)$ and $\alpha_{ij}$ are the coefficients of linear thermal expansion, the strain--energy density gives
\begin{equation}
\psi = \frac12 C_{ijkl}\left[\left(\epsilon_{kl} - \alpha_{kl}(T-T_0)\right)\cdot\left(\epsilon_{ij} - \alpha_{ij}(T-T_0)\right)\right]
\end{equation}
That is $\psi$ is a second order polynomial in $(T-T_0)$
\begin{equation}
\psi = \frac12 C_{ijkl}\epsilon_{kl}\epsilon_{ij} - \left(C_{ijkl}\alpha_{kl}\epsilon_{ij}\right)\cdot (T-T_0)+\left(\frac12 C_{ijkl}\alpha_{kl}\alpha_{ij}\right)\cdot (T-T_0)^2\nonumber
\end{equation}
When only $\epsilon_{12}\neq 0$ with $\gamma=2\epsilon_{12}$, assuming small strains and taking into account the isotropic case, we obtain
\begin{equation}
\psi(\gamma,T) = \frac12 \mu\gamma^2 - 2\gamma\mu\alpha_{12}(T-T_0) + \alpha^2(T-T_0)^2\label{Eq:Psi}
\end{equation}
with $\mu$ the shear modulus (computed and summarized in table~\ref{tbl1}), and $\alpha$ a parameter depending on the elastic moduli and expansion coefficients. Let us now consider thermodynamic variables with $e$ the internal energy and $s$ the entropy per unit volume
\begin{equation}
\psi = e-Ts\label{Psi}
\end{equation}
with
\MS{\begin{equation}
d\psi = de-Tds-sdT = \frac{\partial\psi}{\partial\gamma}\left|_T\right. d\gamma - sdT .\nonumber 
\end{equation}
Since
\begin{equation}
s=-\frac{\partial\psi}{\partial T}\vert_\gamma,
\end{equation}
yielding to
\begin{equation}
\rho\dot{e} = \rho T\dot{s} + \rho\frac{\partial\psi}{\partial\gamma}\dot{\gamma} \label{Eq:de}
\end{equation}
and also}
\begin{eqnarray}
\rho T\dot{s} &=& - \rho T\frac{\partial^2\psi}{\partial t\partial T} = - \rho T\frac{\partial^2\psi}{\partial T^2}\dot{T} - \rho T\frac{\partial^2\psi}{\partial T\partial \gamma}\dot{\gamma} \nonumber\\
&=& \rho C_\gamma\dot{T} - \rho T\frac{\partial^2\psi}{\partial T\partial \gamma}\dot{\gamma}\label{Cgamma}
\end{eqnarray}
with $C_\gamma$ the heat capacity. Considering an elementary volume inside the sample, the heat equation per unit volume, that results from the first principle, is written~\cite{Chrysochoos1989}
\begin{equation}
\rho\dot{e} = \sigma : D + r_{ext} - \text{div}\,{q}\label{Heat}
\end{equation}
with $D$ the strain rates tensor including plasticity, $r_{ext}$ additional heat sources if any, and $\text{div}\,{q}$ standing for heat exchanges at the 
surfaces. Combined with \MS{Eq.~\ref{Eq:de} and} Eq.~\ref{Cgamma}, the heat equation is rewritten:
\begin{equation}
\rho C_\gamma\dot{T} + \text{div}\,{q} = \sigma : D - \rho\frac{\partial\psi}{\partial\gamma}\dot{\gamma} + \rho T\frac{\partial^2\psi}{\partial T\partial \gamma}\dot{\gamma} + r_{ext}\label{Eq:Heat}
\end{equation}
The two first terms \MS{on the right side} account for the intrinsic dissipation rate $\dot\omega_d$ due to mechanical dissipation (plastic deformation)\MS{. We will consider, in the following, that only a part $B$ of the intrinsic dissipation $\dot\omega_d$ is converted into heat, the rest contributing for example to excite coherent waves (emitted noise). The third term in Eq.~\ref{Eq:Heat}} results from the thermo-mechanical couplings. It is called isentropic heat rate $\dot\omega_{is}$ in Ref.~\cite{Chrysochoos1989,Chrysochoos1992}. The last term takes account of eventual external heat sources. At the global scale, a usual approximation proposed by Chrysochoos et al in Ref.~\cite{Chrysochoos1992}, considers that the heat exchanges at interface (heat losses) can be modelled by thermal attenuation with a relaxation time $\tau$ such as 
\begin{equation}
\tau\propto\frac{\rho C_\gamma L_y^2}{\kappa}
\end{equation}
with $\kappa$ the global thermal conductivity. The heat equation is then rewritten \MS{in the absence of external heat sources} 
\begin{equation}
\rho C_\gamma\dot{T} - \frac{\rho C_\gamma}{\tau}T = B\dot\omega_d + \dot\omega_{is}\label{HeatEq}
\end{equation}
with
\begin{equation}
\dot\omega_{is} = -2\mu T\alpha_{12}\dot\gamma
\end{equation}
resulting from Eq.\Ref{Eq:Psi} and with $\dot\omega_d$, the volumic amount of heat given by the conversion of the dissipated mechanical energy (plastic work) into heat
\begin{equation}
\dot\omega_d =\frac{d}{dt}\left\{\int\sigma:d\gamma - \frac12\sigma:\left(\gamma-\gamma_p(\sigma)\right)\right\} 
\end{equation}
\MS{Fig.~\ref{Scheme_Energies_S_XL} (a) illustrates the process of obtaining the plastically dissipated energy, $\omega_d$, from the total mechanical energy stored per unit volume. Assuming a linear elastic material, we compute the plastically dissipated energy by subtracting the reversible elastic contribution. This elastic contribution per unit volume is obtained from the reverse linear curve at each step $\gamma$ (red area in Fig.~\ref{Scheme_Energies_S_XL} (a)). Then, the dissipated energy (green area) is calculated per unit volume, as the difference between the energy introduced into the system and the elastically stored energy. Finally, the derivative with respect to time is taken to obtain $\dot\omega_d$.} To better understand the contribution of the elastic vs. plastic energy in the total mechanical energy stored in the system as a function of time, the figures~\ref{Scheme_Energies_S_XL} (b) and (c) display the different contributions \MS{obtained from the data corresponding to} for the oscillatory shears already discussed and shown in Fig.\ref{Damping_SvsXL}--right ($T_0 = 100~\text{K}$ and $\dot\gamma=10^{10}$~s$^{-1}$) for the S and XL samples. As attempted, the elastic energy per unit volume is restored at the end of each cycle, while the plastically dissipated energy mainly increases. Interestingly, the different \MS{energy densities} stored into the system are not strongly sensitive to the size of the system.

The heat equation~(\ref{HeatEq}) is then solved easily with these well identified contributions. It gives
\begin{equation}
T_{sol}(t) = T_0 e^{t/\tau} + e^{t/\tau}\int_0^t \frac{1}{\rho C_p}e^{-u/\tau} \left(B\dot\omega_d + \dot\omega_{is}\right)~du\label{ResHeatEq}
\end{equation}
The comparison between the numerical solution of the heat equation~(\ref{ResHeatEq}) and the numerical measurement of the \MS{temperature} $T(t)$ allows determining the parameters $\alpha_{12}$ and $B$. For a better comparison with the numerical measurement, a random white noise is also added to the solution $T(t)=T_{sol}(t)+C.N(t)$ with $N(t)\in [-0.5,0.5]$
The best fits obtained are shown in Fig.~\ref{TdeT}. $B$ is chosen smaller than $1$ because only part of the plastically dissipated energy may be converted to heat. The best parameters used are summarized in the Table~\ref{tbl2}. It appears clearly \MS{from Fig.~\ref{TdeT},} that the model is good enough for the highest strain rate. In general, plasticity contributes to increase the global temperature -- sometimes for tens of degrees, while the thermo-mechanical couplings help storing heat into mechanical deformations contributing to global temperature decay. The effective relaxation time $\tau$ appears to depend \MS{also} on the strain rate, with a marked decay at high strain rates that could \MS{eventually} result from the smaller size of atomic clusters involved in the plastic instabilities, and in agreement with the increase of the viscosity with the strain rate~\cite{Fusco2014}. 
The model appears however clearly better for the \MS{large} samples than for the smaller ones. This is not surprising. The small systems are indeed very noisy, \MS{certainly} due to finite size effects in the flowing regime as discussed before. Moreover, \MS{the size dependence observed} in the measured thermal fluctuations visible in the lowest strain rates cases, \MS{could} be related to the noise \MS{observed in these regimes (jerky behaviour of the stress-strain curves)}. 
\begin{table}[h]
\caption{Thermo-mechanical parameters used is Eq.~(\ref{ResHeatEq}) and giving the red fits shown in Fig.~\ref{TdeT}}\label{tbl2}
\begin{tabular*}{\tblwidth}{@{}LLLLLLLLLL@{}}
\toprule
Size & Temperature & Strain rate~[s$^{-1}$] &  $\kappa$ (W.K$^{-1}$.m$^{-1}$) & $C_\gamma$ (J.K$^{-1}$.kg$^{-1}$) & $\tau$ (s) & $\alpha_{12}$ (K$^{-1}$) & $B$ & $C$ (K) & $E$ \\ 
\midrule  
S & $100$ K & $10^{8}$ & $1$ & $327.6$ & $4.07\times10^{-7}$ & $1\times10^{-5}$ & $0$ &  $4$ & $10000$ \\
  &  & $10^{9}$  & $1$ & $327.6$ & $2.04\times10^{-7}$ & $3\times10^{-7}$ & $0.02$ &  $4$ & $5000$ \\ 
  &  & $10^{10}$  & $1$ & $327.6$ & $4.07\times10^{-9}$ & $5\times10^{-6}$ & $0.3$  &  $4$ & $100$ \\ 
\midrule
XL &100 K  & $10^{8}$ & 7 & 327.6 & $1.76\times10^{-7}$ & $1\times10^{-5}$  & $1$ & $0.5$ & 1000 \\
 &  & $10^{9}$ & 7 & 327.6 & $1.76\times10^{-8}$ & $4\times10^{-5}$ & $1$ & $0.5$ & 100 \\
 &  & $10^{10}$ & 7 & 327.6 & $1.76\times10^{-8}$ & $1\times10^{-5}$  & $0.43$ & $0.5$ & 100 \\ 
\bottomrule
\end{tabular*}
\end{table}

\section{Discussion}\label{Discussion}

In this article, we have used \MS{classical} Molecular Dynamics simulations to get insights into the thermo-mechanical behaviour of Cu$_{50}$Zr$_{50}$ metallic glasses. We have shown that plastic deformation induces strong self-heating, possibly reaching few tens of degrees in the samples studied, that is far more than what has been observed in \MS{polycrystals}~\cite{Chrysochoos1992}. This self-heating is generated inside local plastic events at the atomic scale, and it is then transported across the sample. This phenomenon is especially visible in mature (permanent) shear bands - when they exist - that concentrate the plastic activity. \MS{As already observed by Lagogianni et al.~\cite{Lagogianni2022}}, the related temperature rise can reach few percent of the glass transition temperature, with possibly even higher temperatures at the atomic scale. This strong thermal increase is linked to the \MS{classical} kinetic energy acquired during the mechanical instability that occurred at the plastic threshold~\cite{Tanguy2021}. Such numerical evidence of the strong self-heating in metallic glasses is also supported by the experimental observation of structural changes inside shear bands~\cite{Khanouki2023,Deng2023}, that suggests the possibility to overcome locally on a short time the glass transition temperature.

\MS{However, the use of classical simulations does not allow taking into account all phenomena representative of small-scale dissipation in real MG. Especially, the simulations show difficulties in reproducing the heat evacuation mechanisms in a realistic manner in large samples.} To overcome this last problem, we have added a uniform damping parameter inside the system, that represents the general dissipation due to quantum excitations in broken bonds~\cite{Review1983} not considered in the classical description of the interactions. \MS{A more precise approach to thermal dissipation should involve electronic excitations as discussed before, and also electronic temperature and relaxations at short time scales. For that, double-temperature models have been developed~\cite{Ivanov2003, Zhigilei2009, Li2020}, allowing to consider electron-phonon couplings. We consider here, that electronic relaxation is sufficiently fast ($<10^{-13}s$) to be negligible in the process dynamics, but may act indeed as contributing to the global dissipation.}

\MS{We have shown also that }the smallest systems are not representative of the \MS{macroscopic samples} in the plastic regime, because they are unable to support properly \MS{large} shear band formation\MS{, especially in the small strain rate regimes. The shear band width appeared to be at least comprised between $50$~\AA~(for strain localization) and $100$~\AA~(for temperature rises)}. It is thus important to perform simulations on sufficiently large systems \MS{to capture these collective effects. However, }unlike for elastic properties~\cite{Tsamados2009}, it is still an open question to identify the elementary representative volume for shear banding in glasses~\cite{Tanguy2021}. 

\MS{Some characteristic features of the thermo-mechanical behaviour of Metallic Glasses are nonetheless highlighted. We show that the temporality of the self-heating is strain rate dependent. While we observe an excitation followed by a sudden attenuation of the temperature in the form of bursts at a low strain rate ($10^{8}$~s$^{-1}$), it appears in the case of a high strain rate ($10^{10}$~s$^{-1}$), that the self-heating under oscillatory stress manifests itself in the form of loops having a certain temporal persistence. This suggests a reinforced collective thermal behaviour in the high strain-rate case, maybe due to the permanent and somehow more homogeneously spread excitation of the plastic activity in this case~\cite{Fusco2014}. Such a phenomenon is visible only at sufficiently high strain rates, and in large systems.}

The model we have proposed is inspired from classical approaches to thermo-mechanical constitutive laws in solid materials~\cite{Chrysochoos1989}. Three ingredients are taken into account. First, the energy dissipated during plastic deformation acts as a heat source, it means that it is either totally, either partially converted into heat or yields to temperature increase. Second, considering the linear effect of temperature on the linear strain (this effect is equivalent to first order thermal dilatation for volume change but here acting on the shear strain) yields to deformation-driven heat storage, and thus to temperature decrease. Finally, thermal fluctuations, that dominate the thermal response in small systems and in the slow strain rate case, are recovered thanks to the addition of thermal noise. Table~\ref{tbl2} regroups all the parameters of the model obtained at T = $100$~K. It is seen that the part of energy dissipated into plastic deformation and converted into heat source decreases from $100\%$ to $43\%$ for the largest strain rate and is negligeable in the (too) small systems. The measurements in the small systems are not significant, since this size of the system is smaller than the width of the shear band appearing in the large systems. More impressive, within the limitations of our fitting procedure, the relaxation time needed to evacuate heat decreases for orders of magnitude with the strain rate, suggesting that collective effects take place in the high strain rate regime, that increase the efficiency of heat transportation (that is the effective heat conductivity). \MS{Moreover, the linear thermo-mechanical coupling model appears to be quite good at the largest strain rate but differs from the numerical data for smaller strain rates. Especially, the intrinsic dissipation is too small at low strain rates. This term should be increased, may be with higher order sensitivity to the temperature to improve the temperature decay after each peak.} It would be interesting \MS{now} to compare the different parameters of the thermo-mechanical model to experimental observations. To our knowledge, there are unfortunately no \MS{experimental} data available at the moment for the effective thermal conductivity, as well as for the \MS{thermo-mechanical couplings} at different strain rates in these systems. \MS{But independently measured parameters like the thermal expansion~\cite{Gangopadhyay2020} are of the same order of magnitude as those obtained here as a fitting parameter in our model.}

Note that temperature and strain rate play definitely different roles \MS{in this behaviour}. For example, the thermal strain, that is proportional to the temperature, is not very sensitive to the strain rate, while the strain rate decreases the conversion of mechanically dissipated energy into heat but increases the efficiency of heat transportation (parameter $E$). The figures~\ref{SBNADTXL10} and~\ref{SBNADTXL100} show that temperature increases the strain fluctuations at the frontier of the shear bands, while the effect of the shear rate is to split the shear bands into filaments. Finally, the design of the resulting shear band networks appears to depend on the strain rate as well as on the temperature. The effect of temperature and strain rate on the thermal conductivity has \MS{now} to be deepened. Our preliminary measurements of the thermal conductivity inside and outside shear bands has shown an increase inside the shear bands. In the analysis provided here, we do not include the possibility for a progressive change of the parameters as a function of the residual strain. This one is increased during the deformation and certainly depends on the temperature. \MS{It may also affect the amount of plastic work converted into heat~\cite{Ravichandran2002}.} Moreover, using large strain constitutive laws (involving for example second gradients) is also a way of improving our model.

\section{Conclusions}\label{Concluding}

\MS{Using classical molecular dynamics simulations with an effective description of quantum dissipation, we have studied self-heating in a model metallic glass submitted to cyclic loads. The effect of system size, strain rates and external temperatures has been studied in detail. The thermo-mechanical behaviour is compared to a continuum modelization, with linear thermo-mechanical couplings allowing temperature decays. The heating of the glass is due to the conversion of a fraction of plastic work (between $43$\% and $100$\% in the largest most representative system) into heat. The temperature increase is mainly located inside the shear bands. }

\MS{The thermo-mechanical behaviour is shown to display strong finite-size effects, still not converging in  the largest sample studied, whose length is already $0.2 \mu$m. This shows that finite size effects are far more important for shear banding than for the collective elastic behaviour already extensively studied~\cite{Tanguy2002} in amorphous materials. The shear band width, measured from the temperature variation, reaches $100$~\AA~in our sample, in the low strain rate regime. It results from large collective behaviour.}

\MS{This behaviour is also very sensitive to the strain rate, since cooperative effects at a collective scale, associating regular temperature increase, and progressive decay during unloading, appear only at sufficiently large strain rate. For large system sizes however, an additional global effective viscous damping related to quantum sources of relaxation (bonds breaking, or electronic excitations for example) has to be added to the classical description of the forces in order to recover the usual succession of temperature increases and decreases, and embed unphysical uncontrolled heating of the sample. It is thus very important to have a faithful description of the dissipative processes at a microscopic level, to provide realistic simulations. }

\MS{The resulting temperature variations are not very large compared to experimental results for bulk metallic glasses. For ZrCuAgAl and TiZrNiBeCu, for instance, it has been shown recently that the temperature variations around the shear band reach $\Delta T \approx 1200$~K~\cite{Asadi2023}. At ambient temperature, it even becomes possible to exceed the glass transition temperature of Cu$_{50}$Zr$_{50}$, corresponding to $T_g = 670$~K~\cite{Men2005}. At the microscopic scale, we have shown that thermal increase is located mainly inside the shear bands. The shear bands at low strain rate are large and stable, while the alignment of plastic events at large strain rate gives rise to a complex structure for plastic flow, thinner that for smaller strain rates, and with transverse branches. Considering that one promising way of nanostructuration of glasses involves nanocrystalline inclusions~\cite{SepulvedaMacias2018a,Feng2019,Desmarchelier2021}, it would be interesting to now monitor the shape of the shear band network, to generate, without the need of nanocrystals, new materials with well controlled thermo-mechanical properties, and thermal conductivity.}

\section{Acknowledgements}
This work was supported by the RATES project founded by the French National Research Agency (ANR-20-CE08-0022)


\printcredits
\bibliographystyle{unsrt}

\bibliography{biblio}

\end{document}